\documentclass{iopart}

\usepackage{epsfig}
\usepackage{color}
\usepackage{amsfonts}
\usepackage{amsmath}
\usepackage{xypic}
\newcommand{\spacer}{\rule[0cm]{0cm}{0cm}}

\def\inv{^{-1}}
\def\of{\mathbin{\circ}}
\def\mb{$\begin{displaystyle}}
\def\me{\end{displaystyle}$\ }

\newcommand{\R}{{\mathbb R}}

\newcommand{\Proj}{{\mathbb P}}
\newcommand{\C}{{\mathbb C}}
\newcommand{\Quat}{{\mathbb H}}

\newcommand{\Bloch}{\mbox{\rm Bloch}}

\newcommand{\QuatHopf}{\mbox{\rm QuatHopf}}
\newcommand{\OrigHopf}{\mbox{\rm HopfClassic}}
\newcommand{\reverse}{\mbox{\rm reverse}}

\newcommand{\chart}{\mbox{\rm chart}}
\newcommand{\stereo}{\mbox{\rm stereo}}

\begin{document}

\title[Survey of Hopf Fibrations and Rotations]{Survey of Hopf
Fibrations and Rotation Conventions in Mathematics and Physics}

\author{David W. Lyons}
\address{Mathematical Sciences\\
Lebanon Valley College}
\ead{lyons@lvc.edu}

\begin{abstract}
We present a unifying framework for understanding several different
versions of the Hopf fibration, and use this framework to reconcile two
methods of representing rotations of 3-space by unitary matrices---the
mathematician's convention based on quaternion algebra, and the
physicist's convention based on the Bloch sphere.  
\end{abstract}

\maketitle

\section{Introduction}

Sophus Lie made a profound contribution to mathematics and physics in
the late 19th century by developing a theory based on his observation
that solutions to certain problems in mechanics must be invariant under
rigid motions of space, and that the structure in symmetry groups can be
exploited to solve differential equations~\cite{ackerman}.  Although Lie
theory is a rare find in the undergraduate curriculum, one of its
topics---the special orthogonal group $SO(3)$ of rotations of space---is
impossible to miss in courses such as linear algebra, differential
equations and quantum mechanics.

In theory and in practical computations, mathematicians and physicists
use $2\times 2$ unitary matrices as a replacement for $3\times 3$ real
orthogonal matrices.  How this is done, and more important, why this is
natural, are main points of this article.  The explanation rests on the
{\em Hopf fibration}.  Our secondary aim is to reconcile the differences
between math and physics conventions in the use of unitary matrices to
represent rotations.  This is accomplished by comparing different
versions of the Hopf fibration.

The exposition presented here requires no special background beyond
university level vector calculus, linear algebra, and an introduction to
group theory.  Definitions for those few objects which may exceed this
minimum background---projective space, higher dimensional spheres,
commutative diagrams and quaternions---are given in the Appendix.

\section{A Survey of Hopf Fibrations}

Heinz Hopf defined a mapping in his 1931 paper~\cite{hopforig} that we
now call the Hopf fibration.  It was a landmark discovery in the
young subject of algebraic topology that has since been recognized in
many guises in mathematics and physics with applications including
magnetic monopoles, rigid body mechanics, and quantum information
theory~\cite{lyons}. 

The heart of the Hopf map is the canonical projection
\begin{equation}\label{c2top1proj}
\C^2\setminus\{{\mathbf 0}\} \stackrel{\pi}{\to} \Proj^1
\end{equation}
that sends the complex vector $(z,w)$ to its equivalence class $[z,w]$
in projective space.  We interpret this as a map $S^3\to S^2$ by
identifying $S^3$ as the subset of norm 1 vectors in $\R^4=\C^2$, and by
identifying $\Proj^1$ with $S^2$.  The latter identification is a
two-step procedure.  First identify $\Proj^1$ with the extended complex
plane $\C^+=\C\cup \{\infty\}$.  One way to do this is the map
$$
  \chart\colon \Proj^1 \to \C^+
$$
given by $[z_0,z_1] \mapsto z_0/z_1$ (``chart'' is for ``coordinate
chart'').  Second,  identify $\C^+$ with $S^2$ using some version of
stereographic projection.  We will use two stereographic
projections, $\stereo_j\colon S^2 \to \C^+$ for $j=1,3$, given by
\begin{eqnarray*}
  \stereo_1 (x,y,z) &=& \frac{y}{1-x} + i\frac{z}{1-x}\\
  \stereo_3 (x,y,z) &=& \frac{x}{1-z} + i\frac{y}{1-z}
\end{eqnarray*}
and $\stereo_1(1,0,0)=\infty = \stereo_3(0,0,1)$.
We put these
maps together to form a template for the generic Hopf map.  Here and in
diagrams that follow, we highlight the core map~(\ref{c2top1proj}) with
a frame.
\begin{equation}\label{hopftemplatedef}
\xymatrix{
S^3 \ar [rr]^-{\txt{inclusion}} && \C^2\setminus\{{\mathbf 0}\}
\ar[r]^-{\pi} & \Proj^1 \ar[r]^{\chart} & \C^+ \ar[rr]^{\stereo_j\inv}
&& S^2
\save "1,3"."1,4"*[F-]\frm{} \restore
}
\end{equation}
H.~Hopf's original map~\cite{hopforig} arises from this template
by choosing $j=3$.  One obtains variations by altering the
identifications with $S^3$ on the left and with $S^2$ on the right, for
example, by using alternative coordinate charts on $\Proj^1$ and by
choosing different basepoints for stereographic projection.  These
variations are motivated by the desire to adapt coordinates to fit
particular interpretations.

The projection~(\ref{c2top1proj}) comes to life when we view it in terms
of group action.  In general, when a group $G$ acts on a set $X$, we
have a bijection\footnote{Under the right conditions, when $G$ and $X$
are manifolds, the bijection~(\ref{gpactionbijection}) is a
diffeomorphism~\cite{brockertomdieck}.  It is not necessary to know
about manifolds or diffeomorphisms to follow our presentation.}
\begin{equation}\label{gpactionbijection}
  G/I_x \leftrightarrow {\cal O}_x  
\end{equation}
given by $gI_x \leftrightarrow gx$ for each $x\in X$, where $I_x=\{g\in
G\colon gx=x\}$ is the isotropy subgroup for $x$ and ${\cal
O}_x=\{gx\colon g\in G\}$ is the orbit of $x$.  We now apply
this fact twice, where the group is $G=SU(2)$ and the actions arise from
the natural action of $G$ on $\C^2$.  First, let $X$ be the the set of
norm 1 vectors in $\C^2$.  The action on $X$ is transitive (${\cal
O}_x=X$ for all $x$) and the isotropy subgroup of every point is
trivial, so we have the identification
\begin{equation}\label{su2tos3byaction}
%%  SU(2) \stackrel{\widetilde{\rule{.5in}{0in}}}{\longrightarrow} S^3
  SU(2) \widetilde{\longrightarrow} S^3
\end{equation}
given by $g \leftrightarrow
g(1,0)$.  
Second, let $X=\Proj^1$.  The action of $G$ on $X$ is transitive, and the
isotropy subgroup of the point $[1,0]$ is the torus 
$$
  T=\left\{\left[\begin{array}{cc}
e^{i\theta} & 0\\
0 & e^{-i\theta}\end{array}\right]\colon \theta\in \R\right\},
$$
so we have the identification
\begin{equation}\label{su2modttoproj1}
  SU(2)/T \widetilde{\longrightarrow} \Proj^1 
\end{equation}
given by $gT \leftrightarrow
g[1,0]$. 
Now we can rephrase the heart of the Hopf map~(\ref{c2top1proj}) as the
map
\begin{equation}\label{hopfasaction}
  SU(2) \to \Proj^1
\end{equation}
given by $g\mapsto g[1,0]$, where ``rephrase'' means the following
diagram commutes.
$$
%\label{twowaystoseehopf}
\xymatrix{
SU(2) \ar[rr]^{\txt{act on $[1,0]$}} \ar[d]_{\txt{act on $(1,0)$}}
&& \Proj^1 \ar@{=}[d]\\
\C^2\setminus\{{\mathbf 0}\} \ar[rr]^-{\pi}& & \Proj^1
}
$$

Now we are ready to define and compare several versions of the Hopf
fibration in terms of~(\ref{c2top1proj}) and~(\ref{hopfasaction}).
We begin with a Hopf fibration expressed in the language of quaternion
algebra.  

We identify the
quaternions $\Quat$ with $\R^4$ and $\C^2$ via
\begin{equation}\label{quatr4c2}
  x_0+x_1{\mathbf i} + x_2{\mathbf j}+ x_3{\mathbf k} \leftrightarrow
  (x_0,x_1,x_2,x_3) \leftrightarrow (x_0+ix_1,x_2+ix_3)
\end{equation}
and regard $\Quat$ as a real vector space with canonical basis
$\{1,{\mathbf i},{\mathbf j},{\mathbf k}\}$ and also as a complex vector
space with canonical basis $\{1,{\mathbf j}\}$.  We identify $\R^3$ with
the pure quaternions, that is, the subspace of $\R^4$ consisting of
points with zero in the first coordinate.  Under this identification,
the name $p$ for point $p=(x,y,z)$ in $\R^3$ shall also denote the
quaternion $p=x{\mathbf i} + y{\mathbf j}+ z{\mathbf k}$.  We identify
the unit length quaternions with $S^3\subset\R^4$.  The 2-sphere
$S^2\subset\R^3$ is identified with the ``equator'' of $S^3$ which is
the set of unit length pure quaternions.  

The group $SU(2)$ is isomorphic with the group of unit quaternions via
the map
\begin{equation}\label{quatmatident}
\left[\begin{array}{cc}
z & w\\
-\overline{w} & \overline{z}
\end{array}\right] \leftrightarrow z+w{\mathbf j} 
\end{equation}
where $(z,w)$ is a unit length vector in $\C^2$~\cite{brockertomdieck}.
The group of unit quaternions is also naturally identified with $S^3$
via~(\ref{quatr4c2}).  It is important to note that we now have two
distinct identifications of $SU(2)$ with $S^3\subset \C^2$.  The matrix
$\displaystyle \left[\begin{array}{cc} z & w\\ -\overline{w} &
\overline{z}
\end{array}\right] $
identifies with $(z,w)$ by~(\ref{quatmatident}) and identifies with
$(z,-\overline{w})$ by~(\ref{su2tos3byaction}).  We will denote by $T$
the map
$$
  T\colon S^3\stackrel{(\ref{quatmatident})}{\to}  SU(2)
  \stackrel{(\ref{su2tos3byaction})}{\to} S^3
$$
given by $(z,w)\mapsto (z,-\overline{w})$ that arises from combining
these identifications.  We call it $T$ for ``transpose'' because this is
the map you get when you interpret the quaternion as a matrix
by~(\ref{quatmatident}), transpose it, then reinterpret as a point in
$\C^2$ by~(\ref{quatmatident}).  In real coordinates, transpose is given
by $(a,b,c,d)^T= (a,b,-c,d)$.  

The group of unit quaternions acts naturally on the subspace of pure
quaternions (where we interpret the pure quaternions as $\R^3$,
see~\cite{henle} and~\cite{lyons} for details) via
\begin{equation}\label{quatactonr3}
  S^3 \times \R^3 \to \R^3
\end{equation}
given by $(g,p)\mapsto gpg^\ast$, 
where $p$ is a pure quaternion, $g$ is a unit quaternion and $g^\ast$ is
the conjugate of $g$ (the conjugate of $x_0+x_1{\mathbf i} + x_2{\mathbf
j}+ x_3{\mathbf k}$ is $x_0-x_1{\mathbf i} - x_2{\mathbf j}- x_3{\mathbf
k}$ and is what you get if you take the hermitian (conjugate transpose)
of $g$ viewed as a matrix via~(\ref{quatmatident})).  This action
preserves the Euclidean length of $p$, and so restricts to an action on
$S^2$.
\begin{equation}\label{quatactons2}
  S^3 \times S^2 \to S^2
\end{equation}
We choose the basepoint
$p_0=(1,0,0)={\mathbf i}$ and define a map $S^3 \to S^2$ by
\begin{equation}\label{quathopf}
g \mapsto g{\mathbf i}g^\ast.
\end{equation}
The action~(\ref{quatactons2}) is transitive and the isotropy subgroup
of $p_0$ is $\{e^{i\theta}\}$.  As matrices, this isotropy subgroup is
the same as the torus $T$.  Thus the map~(\ref{quathopf}) identifies
with the Hopf fibration~(\ref{hopfasaction}) as shown in the following
commutative diagram.  The correspondence on the bottom row of the
diagram is given by $g[1,0] \leftrightarrow g{\mathbf i}g^\ast$.
$$
\xymatrix{
T \ar@{<->}[r] \ar[d]_{\txt{inclusion}}  &
  \{e^{i\theta}\}\ar[d]^{\txt{inclusion}} \\
SU(2) \ar@{<->}[r]^{(\ref{quatmatident})} \ar[d]_{(\ref{hopfasaction})} & S^3
\ar[d]^{(\ref{quathopf})}\\
\Proj^1 \ar@{<->}[r] & S^2
}
$$

Another Hopf map (although it is rarely if ever identified as such)
arises from a coordinate system on $S^2$ called the Bloch
sphere\footnote{Named after Felix Bloch, recipient of the 1952 Nobel Prize
  in physics.}.  It is
defined as follows: given $(a,b)$ in $\C^2$ with $a$ real, the equations
$a=\cos \theta/2$ and let $b=e^{i\phi}\sin \theta/2$ determine spherical
coordinates $(\theta,\phi)$ for the point 
$$(\cos \phi \sin \theta, \sin \phi \sin \theta,\cos \theta)$$
 on $S^2$.  This is equivalent to the
following.
\begin{equation}\label{blochformula}
\Bloch(a,b) = \stereo_3\inv (\overline{a/b})
\end{equation}
We will take the map ``Bloch'' to be given by~(\ref{blochformula})
whether or not $a$ is real.  Here is a comparison diagram that shows
how the quaternion action and the Bloch coordinate projection fit into
the generic scheme~(\ref{hopftemplatedef}).  From now on, we will use
the labels ``\OrigHopf'', ``$\QuatHopf$'', and ``$\Bloch$'' to refer to
the Hopf's original map, the map~(\ref{quathopf}), and~(\ref{blochformula}), respectively.
%% \begin{equation}\label{comparethreehopfs}
%% \xymatrix{
%%   &&&&&& S^2\\
%% S^3 \ar[d]_T \ar@{.>}[urrrrrr]^{\QuatHopf}&&&& & \C^+\ar[ur]_-{\stereo_{\mathbf i}\inv} & \\
%%  S^3  \ar@{.>}[rrrrrrdd]_{\Bloch} \ar[rr]^-{\txt{inclusion}}& & \C^2\setminus\{{\mathbf 0}\}
%% \ar[r]^-{\pi} & \Proj^1 \ar[r]^{\chart} & \C^+ \ar[ur]^{\cdot i}
%% \ar[dr]^{\txt{conjugation}} \ar[rr]^-{\stereo_{\mathbf k}\inv}& & S^2\\
%%   &&&&& \C^+ \ar[dr]^-{\stereo_{\mathbf k}\inv}& \\
%%   &&&&&& S^2\\
%% }
%% \end{equation}
$$
%\label{comparethreehopfs}
\spacer\hspace{-.3in}\xymatrix{
\OrigHopf &  &
 S^3  \ar@{^{(}->}[r]&  \C^2\setminus\{{\mathbf 0}\}
\ar[r]^-{\pi} & \Proj^1 \ar[r]^{\chart} 
& \C^+ \ar[rrr]^-{\stereo_{\mathbf k}\inv}&& & S^2\\
\QuatHopf & S^3 \ar[r]^-{T} &
 S^3  \ar@{^{(}->}[r]&  \C^2\setminus\{{\mathbf 0}\}
\ar[r]^-{\pi} & \Proj^1 \ar[r]^{\chart} & \C^+ \ar[r]^{\cdot i}
& \C^+ \ar[rr]^-{\stereo_{\mathbf i}\inv}& & S^2\\
\Bloch &  &
 &  \C^2\setminus\{{\mathbf 0}\}
\ar[r]^-{\pi} & \Proj^1 \ar[r]^{\chart} & \C^+ \ar[r]^{\txt{conjugate}}
& \C^+ \ar[rr]^-{\stereo_{\mathbf k}\inv}& & S^2
\save "1,4"."1,5"*[F-]\frm{} \restore
\save "2,4"."2,5"*[F-]\frm{} \restore
\save "3,4"."3,5"*[F-]\frm{} \restore
}
$$
%% (x_1,x_2,x^3,x^4) \ar[r] & (x_1+ix_2,x_2+ix^3) \ar[r] &
%%    [x_1+ix_2,x_2+ix^3] \ar[r] & \frac{x_1+ix_2}{x_2+ix^3} \ar[r] &
%%    (2(x_1x_3 +x_2x_4),2(x_2x_3-x_1x_4),x_1^2 + x_2^2 -x_3^2 -x_4^2)
The following commutative diagram demonstrates identifications among
Hopf fibrations appearing vertically in dotted line frames.  Hopf's original map
is the second column from the left.
$$
%\label{comparehopfs}
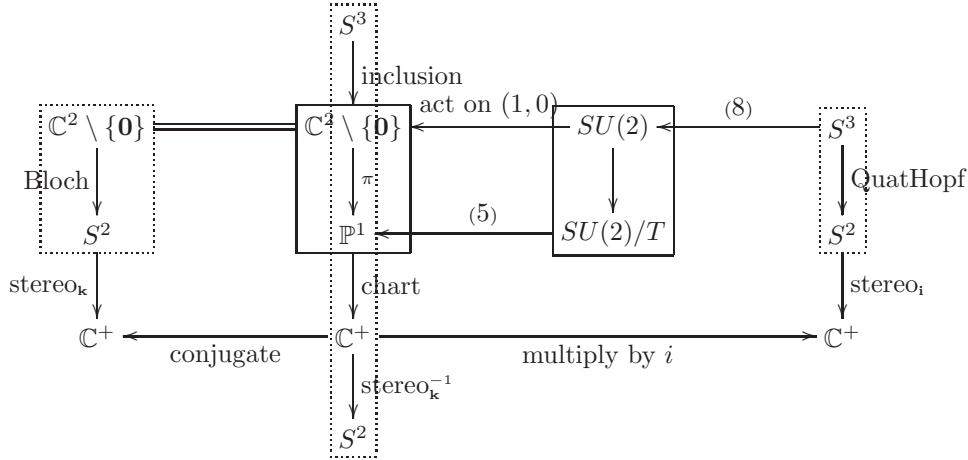
\xymatrix{
&& S^3 \ar[d]^{\txt{inclusion}} &&&&\\
\C^2\setminus\{{\mathbf 0}\} \ar[d]_{\Bloch}\ar@{=}[rr]
&& \C^2\setminus\{{\mathbf 0}\} \ar[d]^{\pi} &&
SU(2) \ar[d] \ar[ll]_-{\txt{act on $(1,0)$}} && S^3 \ar[d]^{\QuatHopf}\ar[ll]_-{(\ref{quatmatident})}\\
S^2 \ar[d]_{\stereo_{\mathbf k}} && \Proj^1 \ar[d]^{\chart} &&
SU(2)/T \ar[ll]_{(\ref{su2modttoproj1})} && S^2 \ar[d]^{\stereo_{\mathbf i}}\\
\C^+ &&  \C^+ \ar[ll]^{\txt{conjugate}} \ar[rrrr]_{\txt{multiply by
    $i$}} \ar[d]^{\stereo_{\mathbf k}\inv}&&  && \C^+ \\
&& S^2 &&&&
\save "2,1"."3,1"*[F.]\frm{} \restore
\save "2,3"."3,3"*[F-]\frm{} \restore
\save "1,3"."5,3"*[F.]\frm{} \restore
\save "2,5"."3,5"*[F-]\frm{} \restore
\save "2,7"."3,7"*[F.]\frm{} \restore
}
$$

We conclude with one more comparison (by commutative diagram) of
$\Bloch$ and $\QuatHopf$.  The label ``$\reverse$'' denotes the
reflection of $\R^3$ that sends $(x,y,z)$ to $(z,y,x)$.
\begin{equation}\label{compareblochhopf}
\spacer\hfill
\xymatrix{
\C^2\setminus\{0\} \ar[d]_{\Bloch} & S^3 \ar[l]_-T \ar[d]^ {\QuatHopf}\\
S^2 & S^2 \ar[l]^{\reverse}
}
\hfill\spacer
\end{equation}

\section{Rotations by Hopf Actions}
%{\bf 3. ROTATIONS BY HOPF ACTIONS}

In the action~(\ref{quatactonr3}) of the unit quaternions on $\R^3$, the
quaternion $g=a + b{\mathbf i} + c{\mathbf j} + d{\mathbf k}$ acts as a
rotation by $\theta/2$ radians about the axis specified by the unit
length vector $\hat{n}=(n_1,n_2,n_3)$ where $\theta,\hat{n}$ are given
by the following equations~\cite{henle,lyons}.
\begin{eqnarray*}
  a &=& \cos \theta/2\\
  (b,c,d) &=& \sin \theta/2 \; \hat{n}
\end{eqnarray*}
Given a real number $\theta$ and a point $\hat{n}$ on
$S^2$, let 
$$
  g_Q = g_Q(\theta,\hat{n}) = \cos \theta/2 + \sin \theta/2 (n_1{\mathbf i} +n_2{\mathbf
  j}+ n_3{\mathbf k}).
$$
We view $g_Q$ both as
a quaternion and as the matrix
$$
 g_Q =
\left[\begin{array}{cc}
\cos \theta/2 + in_1\sin \theta/2 & \sin \theta/2 (n_2 + in_3)\\
\sin \theta/2 (-n_2 + in_3) & \cos \theta/2 - in_1\sin \theta/2
\end{array}\right]
$$
associated via~(\ref{quatmatident}).  Let us denote by
$R(\theta,\hat{n},p)$ the image of $p$ under the rotation by $\theta$
radians about the axis specified by $\hat{n}$.  Then we have
$$
  g_Qpg_Q^\ast = R(\theta,\hat{n},p).
$$
We can also write $R(\theta,\hat{n},p)$ in terms of the Hopf
fibration in the following way.  Let $h_Q$ be any preimage of $p$ under
$\QuatHopf$.  Then we have
$$
  \QuatHopf(g_Qh_Q) = R(\theta,\hat{n},p).
$$
Here is the one-line proof.
$$
  \QuatHopf(g_Qh_Q) = (g_Qh_Q)i(g_Qh_Q)^\ast =
  g_Q(\QuatHopf(h_Q))g_Q^\ast = g_Qpg_Q^\ast.
$$ There is a corresponding expression in terms of
$\Bloch$~\cite{nielsenchuang}.  Given a real number $\theta$ and a point
$\hat{n}$ on $S^2$, let
$$
  g_B = g_B(\theta,\hat{n}) =
\left[\begin{array}{cc}
\cos \theta/2 - in_3\sin \theta/2 & \sin \theta/2 (-n_2 - in_1)\\
\sin \theta/2 (n_2 - in_1) & \cos \theta/2 + in_3\sin \theta/2
\end{array}\right].
$$
Let $h_B$ be
any preimage of $p$ under $\Bloch$.  Then we have
$$
  \Bloch(g_Bh_B) = R(\theta,\hat{n},p).
$$
%% (Note and question: This is exercise~4.6 on p.~175
%% in~\cite{nielsenchuang}.  Is it possible to prove this directly without
%% appealing the the Fubini-Study metric?  It seems that proving that the
%% left-hand side gives an action on $S^2$ by rigid motions is not
%% accessible for sophomore level undergraduate students.)

The purpose of the remainder of this section is to explain the equality
\begin{equation}\label{toshow}
  \QuatHopf(g_Qh_Q) = \Bloch(g_Bh_B).
\end{equation}

First observe that the multiplications $g_Qh_Q$ and $g_Bh_B$ are {\em
  different} operations.  The binary operation in the expression
  $g_Qh_Q$ is quaternion multiplication or matrix multiplication,
  depending on whether you view $g_Q,h_Q$ as quaternions or matrices.
  The binary operation in $g_Bh_B$ is the multiplication of the $2\times
  2$ matrix $g_B$ by the $2\times 1$ vector $h_B$. To keep track of this
  distinction, we will write $g_B\odot h_B$ to denote the latter
  operation.  Having pointed out the difference, we now relate the two
  operations.  Let $\tilde{h_B}$ denote the quaternion associated to
  $h_B$ by~(\ref{quatr4c2}), that is, if $h_B = (z,w)$, then $\tilde{h_B} = z + w{\mathbf
  j}$.  Then we have
\begin{equation}\label{matvectmultasquatmult}
  g_B \odot h_B = \tilde{h_B}g_B^T
\end{equation}
where the operation on the right-hand side is quaternion multiplication and we
view $g_B$ as a quaternion by~(\ref{quatmatident}), or the operation is
matrix multiplication where we view $\tilde{h_B}$ as a $2\times 2$
matrix by~(\ref{quatmatident}).

Now we can derive~(\ref{toshow}).  We have
\begin{eqnarray}
  \Bloch(g_B \odot h_B) &=& \Bloch(\tilde{h_B}g_B^T) \label{line1}\\
&=& \reverse(\QuatHopf(g_B\tilde{h_B}^T)) \label{line2}\\
&=& g_Qpg_Q^\ast.\label{line3}
\end{eqnarray}
The first equality~(\ref{line1}) is by~(\ref{matvectmultasquatmult}).
The second equality~(\ref{line2}) is by~(\ref{compareblochhopf}).  Here
is a geometric explanation for the final equality~(\ref{line3}).
Interpret $h_B^T$ as a $\QuatHopf$ lift of $(z,y,x)$ (by virtue
of~(\ref{compareblochhopf})) and interpret $g_B$ as a (``quat'')
rotation by $-\theta$ around $(n_3,n_2,n_1)$, so
$\QuatHopf(g_B\tilde{h_B}^T)$ calculates
$R(-\theta,\reverse(\hat{n}),\reverse(p))$.  So reversing this result is
the same as rotating $p$ by $g_Q$.  Thus we have completed our goal or
reconciling Bloch sphere rotation conventions with the standard
quaternion approach.  We conclude with a commutative diagram that
expresses~(\ref{toshow}).  
$$
%\label{compareblochhopfrot}
\spacer\hfill
\xyoption{curve}
\xymatrix{
(g_Q,h_Q)\ar@/^1pc/@{.>}[rrrr] & S^3 \times S^3 \ar[rr]_-{\txt{quat. mult.}}
  && S^3\ar[d]^{\QuatHopf} & g_Qh_Q\\
((e^{i\theta},\hat{n}),p) \ar@/^1pc/@{.>}[rrrr] \ar@{.>}[u]\ar@{.>}[d]& (S^1\times S^2)\times S^2 \ar[rr] \ar[u]
  \ar[d]&& S^2 & R(\theta,\hat{n},p)\\
(g_B,h_B) \ar@/_1pc/@{.>}[rrrr]& SU(2)\times \C^2 \ar[rr]^-{\odot} && \C^2 \ar[u]_{\Bloch} &
  g_B\odot h_B
}
\hfill\spacer
$$

%% \xyoption{curve}
%% \xymatrix{
%%  U \ar@/_/[ddr]_y \ar[dr] \ar@/^/[drr]^x \\
%%   & X \times_Z Y \ar[d]^q \ar[r]_p
%%                  & X \ar[d]_f            \\
%%   & Y \ar[r]^g   &Z                      }

\section{Appendix}

The set $\Proj^1$, called {\em 1-dimensional complex projective space},
is the set of equivalence classes in $\C^2\setminus\{{\bf 0}\}$, where
${\bf 0}$ denotes the zero vector ${\bf 0}=(0,0)$, with respect to the
equivalence relation $\sim$ defined by $(z,w)\sim (z',w')$ if and only
if $(z,w)=\lambda (z',w')$ for some nonzero complex scalar $\lambda$.

The set $S^n$, called the {\em $n$-dimensional sphere}, or simply the
{\em $n$-sphere}, is the set of points $(x_0,x_1,\ldots,x_n)$ in
$\R^{n+1}$ that satisfy
$$x_0^2 + x_1^2 + \cdots + x_n^2 = 1.$$

To say that a diagram of sets and functions
  {\em commutes} means that if there are two different function
  compositions that start at set $A$ and end at set $B$, then those
  compositions must be equal as functions.  For example, to say the
  following diagram commutes means that $r\of t = b\of \ell$.
$$
\xymatrix{
A \ar[r]^{t} \ar[d]_{\ell}
& X \ar[d]^{r}\\
Y \ar[r]^{b}&  B
}
$$

The {\em quaternions} are the set $\R^4$ endowed with a noncommutative
multiplication operation, given below.  The standard basis vectors
$$(1,0,0,0),(0,1,0,0), (0,0,1,0), (0,0,0,1)$$ are denoted $1,{\mathbf
i},{\mathbf j},{\mathbf k}$, respectively, so that the vector
$(x_0,x_1,x_2,x_3)$ in $\R^4$ is written $x_0 + x_1{\mathbf i} +
x_2{\mathbf j}+ x_3{\mathbf k}$ as a quaternion.  The multiplication is
determined by the relations
$${\mathbf i}^2={\mathbf j}^2={\mathbf k}^2=-1$$
$${\mathbf i}{\mathbf j}={\mathbf k} \hspace{.25in} {\mathbf j}{\mathbf
  k}={\mathbf i} \hspace{.25in}{\mathbf  k}{\mathbf i}={\mathbf j}$$
$${\mathbf j}{\mathbf i}=-{\mathbf k} \hspace{.25in} {\mathbf k}{\mathbf
  j}=-{\mathbf i} \hspace{.25in} {\mathbf i}{\mathbf k}=-{\mathbf j}$$
and extending linearly.


\medskip

\begin{thebibliography}{10}

\bibitem{ackerman}
M.~Ackerman et al., editors.
\newblock {\em Lie Groups: History, Frontiers and Applications, Vol. 1.}
\newblock Math Sci Press, Brookline, MA, 1975.

\bibitem{hopforig}
H.~Hopf.
\newblock \"Uber die Abbildungen der dreidimensionalen Sph\"are auf die
Kugelfl\"ache.
\newblock {\it Math. Ann.}~104:637--665, 1931.

\bibitem{brockertomdieck}
Theodor Br\"ocker and Tammo tom Dieck.
\newblock Representations of Compact Lie Groups.
\newblock Springer-Verlag, 1985.

\bibitem{nielsenchuang}
Michael~A. Nielsen and Isaac~L. Chuang.
\newblock {\em Quantum Computation and Quantum Information}.
\newblock Cambridge University Press, 2000.

\bibitem{henle}
Michael Henle.
\newblock {\em Modern Geometries, 2nd edition}.
\newblock Prentice Hall, 2001.

\bibitem{lyons}
David~W. Lyons.
\newblock An Elementary Introduction to the Hopf Fibration.
\newblock {\em Mathematics Magazine}~76(2):87--98, 2003. 


\end{thebibliography}
\end{document}